\documentclass[aps,prl,showpacs,twocolumn,superscriptaddress]{revtex4}
\usepackage{graphicx}

\begin{document}
\bibliographystyle{prsty}

\title{Wealth redistribution with  finite resources}

\date{\today}

\author{S. Pianegonda}
\email{salete@if.ufrgs.br}
\affiliation{Instituto de F\'{\i}sica, Universidade Federal do Rio Grande do Sul,
C.P. 15051, 91501-970 Porto Alegre, Brazil}

\author{J.R. Iglesias}
\email{iglesias@lps.u-psud.fr}
\affiliation{Instituto de F\'{\i}sica, Universidade Federal do Rio Grande do Sul,
C.P. 15051, 91501-970 Porto Alegre, Brazil}
\affiliation{Laboratoire de Physique des Solides, Universit\'{e}
Paris-Sud, B\^atiment 510, 91405 Orsay, France}

\author{G. Abramson}
\affiliation{Centro At\'{o}mico Bariloche, Instituto Balseiro and CONICET, 8400 San
Carlos de Bariloche, Argentina}

\author{J.L. Vega}
\affiliation{Banco Bilbao Vizcaya Argentaria, Via de los Poblados s/n, 28033
Madrid, Spain}

\begin{abstract}
We present a simplified model for the exploitation of finite resources by
interacting agents, where each agent receives  a random fraction of the
available resources. An extremal dynamics ensures that the poorest agent has a
chance to change its economic welfare. After a long transient, the system
self-organizes into a critical state that maximizes the average performance of
each participant. Our model exhibits a new kind of wealth condensation, where
very few extremely rich agents are stable in time and the rest stays in the
middle class.

\end{abstract}

\pacs{87.23.Ge, 89.65.Gh, 89.75.Da, 45.70.Ht, 05.65.+b}

\maketitle

Extended systems showing critical behavior do not need any fine-tuning of a
parameter to be in a critical state. In an attempt to explain this behavior,
Bak, Tang and Wiesenfeld introduced the concept of self--organized criticality
(SOC) \cite{BTW87}. In the critical state, there are long-range interactions,
by which each part of the system feels the influence of all the others. More
precisely, this means that many of the relevant observables in the system
follow a power--law or Pareto--L{\'e}vy distribution with a non-trivial
exponent.

Economics is, by far, one of the more complex extended systems. Economic
development has always been considered the driving (or relevant) force in
determining the relationships inside a society. Similar to what happens in
paleontology \cite{Raup86-EG72-GE77-EG88}, it follows a {\it punctuated}
pattern: Wars, famines, revolutions (and counter-revolutions) are the most
evident (and extreme) illustrations of these bursts of historical activity. It
is then natural, if nothing else by the force of mere analogy, to look for
evidences of critical behavior in economic systems.

In this paper we will concentrate on one particular aspect of economic
processes. In recent years a great deal of effort has been devoted  to the
analysis of economic data. From stock--exchange fluctuations
\cite{LM99-MS95-Mantegna91}, models of production \cite{BCSW93},  size
distribution of companies \cite{Stanley_ea96}, to the appearance of money
\cite{Donan2000}, and the effects of controls on the market \cite{CVV01}, it
has finally been shown that market economy exhibits properties characteristic
of a critical system \cite{Mandelbrot97-MS97-MS99}.

Here we aim at modeling the competition among different agents (countries,
enterprises, etc.) acting in an environment with constant resources. For this
reason, we call the present model TWC-model (Total Wealth Conserved model).
This restriction has several motivations. On the one hand, it can be argued
that our planet is finite and consequently the resources in it are finite. Even
though there are resources that are actually renewable, we assume that those
are renewed at the expense of others, thus making the totality of available
resources constant. On the other hand any study of wealth increase requires
understanding the behavior of the reference (conservative) system.

We will model our {\it economy} as a one-dimensional lattice, every site of which
represents an agent.  A\-gents with closer ties to each other (geographical or
otherwise) will be neighbors on the lattice. For simplicity sake we assume
periodic boundary conditions. Each agent will be characterized by some
wealth-parameter that represents its welfare. The exact choice of this
parameter is not straightforward. For instance, if we are thinking of countries
in the world economy the GDP, GNP or some function of macroeconomic indicators
could be a reasonable choice. In the case of companies, equity, share price or
some combination of them with outstanding debt are reasonable candidates.
We choose an initial configuration where the wealth is distributed
randomly among agents, the wealth of each agent being between 0 and 1.

In the marketplace, all agents strive to improve their situation. In particular
the poorest agent is the one feeling the strongest pressure to move up the
ladder. Thus, we model this process by an extremal dynamics. At each time step,
the poorest country, i.e. the one with the minimum wealth, will take some
action to improve its economic state. That is, it will change its production
methods, borrow money, increase the percentage of sown fields or take some
other measure aiming at increasing its wealth. Since the outcome of any such
measure is uncertain, we model this outcome as a random change in the wealth
parameter of this country. Moreover, whatever wealth is gained (lost) by the
poorest agent will be at the expense of its neighbors and  we assume it to be
equally divided among its two nearest neighbors. We would like to remark that,
apart from conservation, we do not impose any limit on the wealth evolution, so
negative values are possible, corresponding to agents having debt rather than
wealth. Since a site with negative wealth will most probably be the minimum in
the near future, we expect such a site to linger only  a few steps {\it in
red}. In this simplified version of the model, default is not taken into
account, that is, any company may  stay for ever in debt, albeit with a very
low probability.

\begin{figure}[h]
\centering
\resizebox{\columnwidth}{!}{\includegraphics{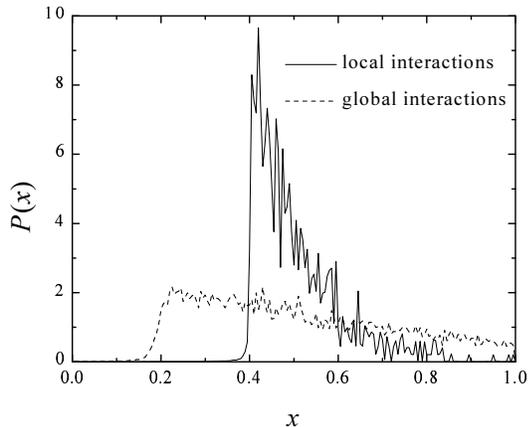}}
\caption{Distribution of wealth. The plots show the probability distribution $P(x)$
for the model with interactions to nearest neighbors (full line) and with
global interactions (dashed line). In both cases the number of agents is
$N=1000$. The histograms were built using $10^5$ consecutive states, after a
transient of $10^6$ steps has elapsed.}
\label{distrisal}
\end{figure}

After a relatively long transient the system arrives at a stationary wealth
distribution; one typical image of the wealth's landscape is shown in
Fig.~\ref{distrisal}.

As in other extremal dynamics models the system self-organizes into a state in
which almost all agents are beyond a certain threshold, $\eta_T \approx 0.4$.
Above threshold, the distribution of agents is exponential, i.e. there are
exponentially few rich agents while the mass of them remain in what we call a
{\it middle class}. Wealth redistribution is then evident.

Turning our attention back to Fig.~\ref{distrisal}, we show also the globally
coupled (mean field) solution,   corresponding  to a random choice of  sites
from which wealth is taken or given to. This mean field solution exhibits a
lower threshold and, more strikingly, an almost linear behavior   beyond
threshold. This departs from standard extremal dynamics models where both
distributions are rather uniform. Furthermore, the distribution of avalanches
follows a power law with the same exponent as the Bak-Sneppen universality
class.

\begin{figure}[h]
\centering
\resizebox{\columnwidth}{!}{\includegraphics{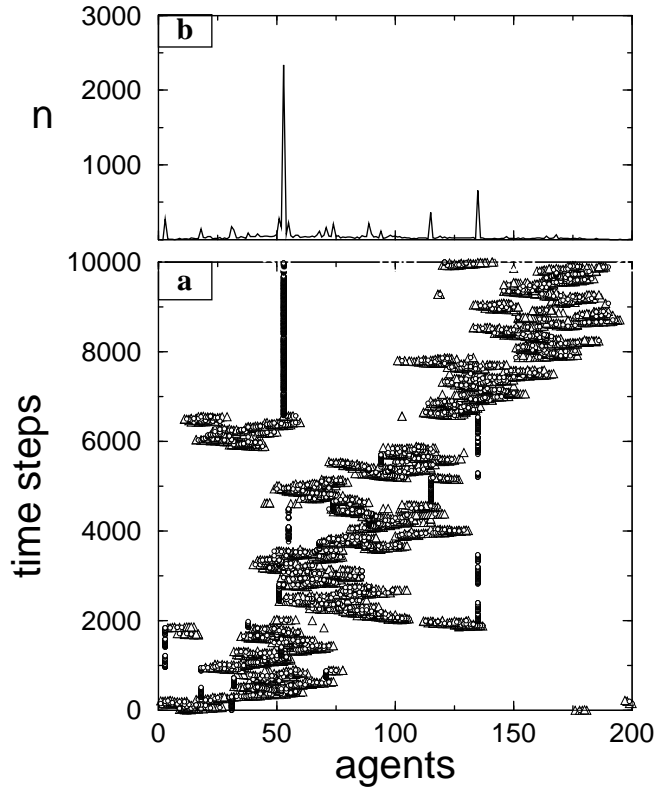}}
\caption{(a) Time evolution of the positions of the maximum (circles) and minimum
(triangles) wealth. (b) Number of time steps, $n$, that a given agent has had
the maximum wealth. The results presented in this figure correspond to $10^4$
time steps after a transient of $2\times10^5$ time steps.}
\label{sem2}
\end{figure}

In Fig.~\ref{sem2}a we show the temporal evolution, in the SOC state, of  the
position of  the systems minimum and maximum wealth. We can see that, while the
site of minimum wealth is changing continuously, generating avalanches of
wealth redistribution  among  neighbors, the richest site is stable over long
periods of time. Indeed, when affected by an avalanche it can recover its
status after a short time. These brief  interruptions, usually produced by
short-lived avalanches,  are reflected as  gaps in the maxima lines.

In Fig.~\ref{sem2}b  we present the statistics of the number of time-steps a
site spends as absolute maximum. Clearly only a few agents have spent most of
the time as maxima, while the rest lurks somewhere in the middle class. We have
also observed that not only the absolute maximum is stable, but also a
privileged group, whose wealth is around the same value of the maximum, remains
in its prosperous position for quite a while. The composition and hierarchy of
this privileged group is barely affected by the avalanches that produce the
abovementioned gaps.

\begin{figure}[h]
\centering
\resizebox{\columnwidth}{!}{\includegraphics{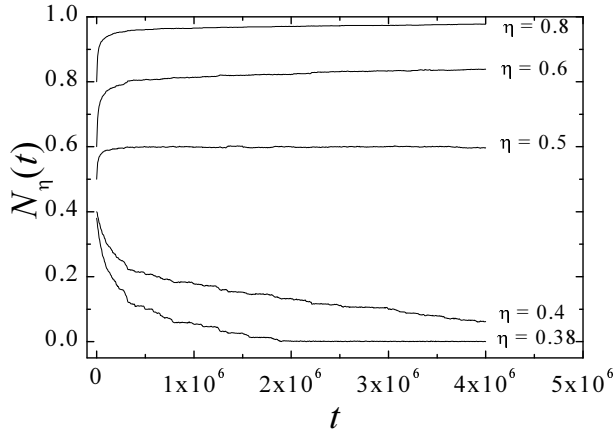}}
\caption{The fraction of the system whose state is below $x=\eta$, $N_\eta(t)$, is
displayed as a function of time, for selected values of $\eta$, as shown in the
legend. In all cases $N=10000$.}
\label{progre3}
\end{figure}

So far we have focused our attention on the final state of the economy, that
is, on the wealth distribution in the self-organized state.  Let us now devote
some time to discuss the transient. In particular, we are interested in
understanding the process of wealth accumulation. In Fig.~\ref{progre3} we show
the time evolution of $N_\eta$, the fraction of the agents whose wealth $x(t) <
\eta$, for different values of $\eta$. Slowly but steadily, for values of $\eta
\leq \eta_T$  these fractions decrease, thus showing the speed of  wealth
redistribution in the system. As expected, the higher the value of  $\eta$, the
slower the progress. As can be clearly seen in the picture, all fractions with
$\eta \leq \eta_T$ converge to zero, while for $\eta > \eta_T$  the fraction
grows quickly to its asymptotic value. When the value of $\eta$ is near 1, the
fraction quickly converges to 1, reflecting the existence of small privileged
groups. We have also observed that the probability of one agent becoming
wealthier in a time step decreases as time goes by, to finally converge to a
finite value, $p\approx 0.76$. Both effects are a consequence of the fact that
the total wealth to be distributed is finite.

 From these results several conclusions may be drawn. First and foremost,
resource conservation leads to an exponential wealth distribution, where the
very few extremely rich agents are stable in time and the rest is just above
threshold. This is tantamount to saying that the invisible hand \cite{Adam} of
redistribution works  only among the middle class. Neither trade nor cost of
debt, returns or tax on wealth is  explicitly included in this model. This
reinforces the role played by geography. As can be seen in
Fig.~\ref{distrisal}, the a-geographic mean field solution generates a
completely different wealth distribution.  Indeed, this globally coupled
solution can be compared with the results obtained in
Refs.~\cite{BM2000,Burda2001} for stochastic multiplicative market models, and
reinforces the conclusions presented in Ref.~\cite{HS2001} concerning wealth
condensation with a finite number of agents. Secondly, the economic progress in
society is steady, even if slow.

At this point it is instructive to compare these results with  Pareto's law
\cite{pareto}, which suggests that individual wealth follows a power law
distribution. Our model exhibits exponential distribution in the local limit
and a particular power law distribution (with an exponent close to one) in the
mean field  limit. As explained above, the later case corresponds to global
interactions, and in this case there is also a higher number of poorer agents,
because the threshold is much lower than in the local interaction case. In
brief, power laws seem to be a consequence of globalization in the market and
favor a wide spectrum of wealth distribution, i.e. increase inequalities. In a
sense, our local model corresponds to a kind of feudal world, where local
barons maintain their dominance for long periods of time.

Summarizing, the model presented here provides a simple description of wealth
redistribution in the early stages of human economic history, and indicates
some of the possible driving forces beyond the market expansions that
influenced this redistribution process.  We believe these conclusions may be of
interest in view of the present debate over the goods and evils of
globalization.

J.R.I. and S.P. acknowledge support from Conselho Nacional de Desenvolvimento
Cient\'{\i}fico e Tecnol\'{o}gico (CNPq, Brazil); J.R.I also acknowledges the
hospitality and support of Universit\'{e} Paris-Sud, Orsay, France, and Facultad de
Ciencias, Universidad de Cantabria, Santander, Spain. We acknowledge partial
support from SETCYP (Argentina) and CAPES (Brazil) through the
Argentine-Brazilian Cooperation Agreement BR 18/00, and G.A. thanks the
hospitality of the Instituto de F\'{\i}sica, Universidade Federal do Rio Grande do
Sul, Porto Alegre, Brazil.

\end{document}